\documentclass[letterpaper, 10 pt, conference]{ieeeconf}  

\IEEEoverridecommandlockouts                              

\newcommand{\st}[1]{${#1}^{\textrm{st}}$}
\newcommand{\nd}[1]{${#1}^{\textrm{nd}}$}
\newcommand{\rd}[1]{${#1}^{\textrm{rd}}$}

\def\C{{\setbox0=\hbox{$\displaystyle{\textrm{C}}$}
		\hbox{\hbox to0pt{\kern 0.4\wd0\vrule height 0.95\ht0\hss}\box0}}}

\def\cF{\mathcal{F}}

\usepackage{color}
\usepackage{xcolor}
\usepackage{colortbl}
\newtheorem{remark}{Remark}

\usepackage{amsmath,amssymb}
\usepackage{rotating}
\usepackage{epstopdf}
\usepackage{tikz} 
\usepackage[siunitx]{circuitikz}
\usepackage[percent]{overpic}
\usepackage{graphicx}

\usepackage{float}
\usepackage{longtable}
\usepackage{diagbox}
\usepackage{graphicx}
\usepackage{subcaption}
\usepackage{xcolor}
\usepackage{pict2e}

 \usepackage{flushend}

\usepackage{epsfig}
\usepackage[normalem]{ulem}

\def\C{{\setbox0=\hbox{$\displaystyle{\textrm{C}}$}
		\hbox{\hbox to0pt{\kern 0.4\wd0\vrule height 0.95\ht0\hss}\box0}}}

\def\cF{\mathcal{F}}

\usepackage{placeins}

\title{{\tt \scriptsize European Control Conference (ECC) --- July 7-10, 2026, Reykjavík, Iceland} \\ \LARGE \bf
Air Supply Control for Proton Exchange Membrane Fuel Cells \\
without Explicit Modeling}

\author{Meziane Ait Ziane$^{1}$, Michel Zasadzinski$^{2}$,  Cédric Join$^{2}$ and Michel Fliess$^{3}$ 
\thanks{*This work was not supported by any organization}
\thanks{$^{1}$M. Ait Ziane is with Université de Lorraine, GREEN, F-54000 Nancy, France.
        {\tt\footnotesize meziane.ait-ziane@univ-lorraine.fr}}%
\thanks{$^{2}$M. Zasadzinski and C. Join are with Research Center for Automatic Control of Nancy (CRAN), UMR CNRS 7039, Université de Lorraine, F-54000 Nancy, France
        {\tt\footnotesize \{michel.zasadzinski,cedric.join\}@univ-lorraine.fr}}%
 \thanks{$^{3}$M. Fliess is with Laboratoire  Jacques-Louis Lions (LJLL), UMR CNRS 7598, Sorbonne Université, 75005 Paris, France
        {\tt\footnotesize michel.fliess@sorbonne-universite.fr, michel.fliess@swissknife.tech }}%
}

\begin{document}

\maketitle
\thispagestyle{empty}
\pagestyle{empty}

\begin{abstract}
\color{black} 
 Our objective is to study the performance and robustness of the model-free strategy for controlling the oxygen stoichiometry of a fuel cell air supply system with a proton exchange membrane. After reviewing the literature on modeling and control of this process, the model-free approach appears to be a good candidate because, on the one hand, it allows straightforward real-time adaptation to track operating points and, on the other hand, it requires a low computational burden, which is attractive for industrial applications.  
Numerical simulations for two scenarios (constant and variable oxygen stoichiometry) with two current profiles reveal satisfactory performance of the model-free control law. The robustness is addressed by considering significant variations in the parameters of the proton exchange membrane air supply system.

\end{abstract}

\section{INTRODUCTION}\label{sec_intro} 
\color{black}

One potential technological solution for decarbonizing the transportation sector is to replace combustion engines with electrical motors supplied by fuel cells. Among the various existing technologies, the {\em Proton Exchange Membrane Fuel Cell} ({\em PEMFC}) is particularly well suited to automotive applications, as it operates at low temperatures (60 to 80°C) and is resilient to load variations   \cite{YLPRJH:21,AYMZD:22}.  

A PEMFC is viewed as a complex system due to the interactions of several physical phenomena (electrical, thermal, chemical, \ldots). It is mandatory to feed the PEMFC with a sufficient amount of hydrogen and oxygen to ensure the desired electrical power and to avoid performance degradation. Thus, air supply control is one of the key points in the proper operation of a PEMFC. Since air is supplied by a compressor, both the fuel cell and the compressor (referred to as the PEMFC air-feed system in the literature) must be taken into account \cite{CGFTKLCW:24}.  




The behavior of the PEMFC system is highly nonlinear. A PEMFC model including the compressor and oriented control has been initially proposed in \cite{PSP:04}. From this model, an air-feed one with 5 states is presented in \cite{LGSWW:19}, with 4 states in \cite{Suh:05b}, and with 3 states in \cite{ZZLP:22}.  These models have been  used widely in the literature for control purposes. The design of all control laws cited in this paragraph is explicitly based on the aforementioned PEMFC air-feed models.



A sliding mode control approach is designed in \cite{MLJW:13} with state feedback and in \cite{ZZLP:22,LHAC:15} by using an observer. A controller based on state feedback linearization is proposed in \cite{CLWOS:18} and in \cite{LYDZGC:22} by adding an extended state observer. A LQR tracking controller based on an observer is synthesized in \cite{LLHG:18}. A control law strategy constituted by three parts, called a triple-step controller, is developed in \cite{MZGCS:20} and \cite{ZGKLH:24}: \st{1} part is a PID controller, \nd{2} part is a steady-state feedforward controller, and \rd{3} is a steady-state feedback.   It is challenging to obtain a model that accurately describes the actual behavior of the PEMFC air-feed system, as the PEMFC stack is difficult to model and to obtain the exact values of the physical parameters. The control laws mentioned in this paragraph are promising strategies, but they have a limitation on their efficiency due to the quality of the model used.

PID are widely proposed in the literature to control PEMFC air-feed system. PID controllers without gain adaptation are proposed:  feedforward PID in \cite{BCLTYS:25},  robust PID in \cite{WaK:10} and PID based on \st{1} order approximation  in \cite{DSD:19}. The gains of PID controllers are adapted by using a fuzzy logic approach in \cite{Ali:18} and \cite{OWLSX:15}, while this adaptation is made using neural techniques in  \cite{LiY:21b}, \cite{YGOK:24}, and \cite{DBLDG:14}. Both fuzzy logic and neuronal approaches are used to adapt PID gains in \cite{GBK:19}, \cite{ZLLXX:21} and \cite{YCH:23}. Considering the high nonlinearity of the model, PID parameter tuning seems required to control the PEMFC air-feed system over a wide operating range.

An artificial intelligence approach is also applied to design a controller for PEMFC air-feed systems without including PID in \cite{WaW:23}. An adaptive B-spline neural network is applied in \cite{SBAR:14}, an adaptive observer is added to the neural network in \cite{WWZX:21}, and distributed deep reinforcement learning is employed in \cite{LiY:21}. For real-time implementation considerations, the use of fuzzy logic and/or neural network techniques to perform either gains adaptation of PID controllers or the design of artificial intelligence controllers  is  computationally expensive. 

The {\em Model-Free Control} ({\em MFC}) developed in \cite{FlJ:13,FlJ:22} can be considered an interesting alternative to the approaches mentioned above: the mathematical model is not necessary to design the control law, which is adapted using an ultra-local model in a straightforward manner.  Furthermore, its implementation is simpler than those mentioned above and requires very few parameters. The aim of this paper is to investigate the performance and robustness of the MFC strategy on a PEMFC air-feed system model under several scenarios.

The outline of this paper is as follows. MFC design is briefly recalled in Sect. \ref{sec_MFC}. The PEMFC air-feed system model used to test the designed controller in this paper and the control objective are given in Section \ref{sec_PEMFC}. The simulation results are presented in Sect. \ref{sec_Sim}: the two scenarios considered in this paper are described in Sect. \ref{ssec_scenarii}; the simulations for both constant and variable desired oxygen stoichiometry, along with robustness analysis, are conducted in Sect. \ref{ssec_sim_constant} and \ref{ssec_sim_variable}, respectively. A conclusion with discussion is given in Sect.~\ref{sec_Conc}. 


\section{SHORT RECALL ON MODEL-FREE CONTROL} \label{sec_MFC}

We only deal here with SISO systems. Elementary functional analysis and differential algebra show \cite{FlJ:13}, as well as concrete experiments, that most, or at least many, systems encountered in practice can be approximated by the \emph{ultra-local} model
\begin{equation}
\dot{y} = \cF + \alpha u
\label{ul}
\end{equation}
where the control and output variables are respectively $u$ and $y$, $\cF$ is a quantity that subsumes the poorly known system structure and the disturbances. The constant $\alpha \in \mathbb{R}$ is chosen by the practitioner such that the three terms in  \eqref{ul} are of the same magnitude: $\alpha$ does not need to be precisely estimated.
The following data-driven real-time estimate $\cF_{\text{est}}(t)$ of $\cF(t)$ is given by 
\begin{equation*}\label{integral}
{\small \cF_{\text{est}}(t)  =-\frac{6}{\tau^3}\int_{t-\tau}^t \left\lbrack (\tau -2\sigma)y(\sigma)+\alpha\sigma(\tau -\sigma)u(\sigma) \right\rbrack d\sigma }
\end{equation*}

Close the loop with the following \emph{intelligent proportional controller}, or \emph{iP},
\begin{equation}\label{ip}
u = - \frac{\cF_{\text{est}} - \dot{y}^\ast + K_p e}{\alpha}
\end{equation}
where $y^\ast$ is the reference trajectory, $e = y - y^\ast$ is the tracking error, $K_p$ is a tuning gain.
Equations \eqref{ul} and \eqref{ip} yield
\begin{equation*}
  \dot{e} + K_p e = \cF - \cF_{\text{est}}.  
\end{equation*}

If the estimate $\cF_{\text{est}}$ is ``good'', \textit{i.e.}, $\cF - \cF_{\text{est}} \approx 0$, then $\lim_{t \to +\infty} e(t) \approx 0$ if and only if, $K_p > 0$. This local stability property proves that the tuning of $K_p$ is straightforward. This is a major difference with the classic gain tuning for PIs and PIDs (see, \textit{e.g.}, \cite{FlJ:22}).

\section{ PEMFC AIR FEED SYSTEM MODELING AND CONTROL OBJECTIVE} \label{sec_PEMFC}

\subsection{PEMFC air-feed system model} \label{ssec_PEMFC_mod}
A PEMFC model oriented control has been initially proposed in \cite{PSP:04}. From this model, Suh \cite{Suh:05b} has proposed an air-feed model, with 4 states, has been used widely in the literature: in  \cite{MLJW:13} and \cite{LHAC:15} for sliding mode control, in \cite{NZO:25} for fault detection with observer and in  \cite{WWZX:21} for both control and state estimation using neural networks. 

This model has been validated on test bench in \cite{SuS:07} and \cite{LLAHW:15} and is described by
\begin{subequations}   \label{eq_model}
\begin{align}
    \dot{x}_1 &= c_1(x_4-x_1-x_2-c_2)-c_7\xi  \notag \\ 
    & \qquad -\dfrac{c_3x_1}{c_4x_1+c_5x_2+c_6} c_{17} \sqrt{x_1+x_2+c_2-c_{11}},  \\
    \dot{x}_2 & = c_8(x_4-x_1-x_2-c_2) \notag \\
    & \qquad -\dfrac{c_3x_2}{c_4x_1+c_5x_2+c_6}c_{17}\sqrt{x_1+x_2+c_2-c_{11}}, \\ 
    \dot{x}_3&= -c_9x_3-c_{10}\left[ \left( \frac{x_4}{c_{11}} \right)^{c_{12}} -1\right] + c_{13} u,  \\ 
    \dot{x}_4 &= c_{14}\left[ 1 + c_{15}\left[ \left( \frac{x_4}{c_{11}}\right)^{c_{12}} -1 \right] \right] \notag \\ 
    & \qquad \qquad \quad \times[ c_{21}x_3 - c_{16}(x_4-x_1-x_2-c_2)],
\end{align}
\end{subequations}
where 
\begin{itemize}
    \item 
      $x_1 = P_{O_2}$ is the oxygen partial pressure,  $x_2 = P_{N_2}$ is the nitrogen partial pressure,  $x_3 = \omega_{cp}$ is the compressor angular speed and  $x_4 = P_{sm}$ is the supplied manifold pressure, 
    \item $\xi$ is the fuel stack current which is considered as measured disturbance in the sequel, 
    \item $u$ is the motor current  and is the control input, 
    \item $c_{i}$ with $i=1,\ldots, 21$ are PEMFC and air-feed compressor parameters given in Tab. \ref{tab_c}, where physical parameters values are given in Table 2 in \cite{MLJW:13}.
\end{itemize}

\begin{table}[h!]
\caption{Constants $c_1$ to $c_{21}$}
\label{tab_c}
\centering
\begin{tabular}{|c||c|}
\hline
$c_1 = \dfrac{RT_{fc} k_{ca,in}x_{O_2}}{V_{ca} M_{O_2}(1 + \omega_{atm})}$ 
& $c_2 = p_{sat}$ \\
\hline
$c_3 = \dfrac{RT_{fc}}{V_{ca}}$ 
& $c_4 = M_{O_2}$ \\
\hline
$c_5 = M_{N_2}$ 
& $c_6 = M_v p_{sat}$ \\
\hline
$c_7 = \dfrac{RT_{fc} k_{ca,in}}{V_{ca} 4F}$ 
& $c_8 = \dfrac{RT_{fc} k_{ca,in}(1 - x_{O_2})}{V_{ca} M_{N_2}(1 + \omega_{atm})}$ \\
\hline
$c_9 = \dfrac{f}{J_{cp}}$ 
& $c_{10} = \dfrac{\eta_{v-c} V_{cpr/tr} \rho_a C_p T_{atm}}{2\pi J_{cp} \eta_{cp}}$ \\
\hline
$c_{11} = p_{atm}$ 
& $c_{12} = \dfrac{\gamma - 1}{\gamma}$ \\
\hline
$c_{13} = \dfrac{\eta_{cm} k_t}{J_{cp}}$ 
& $c_{14} = \dfrac{RT_{atm}}{M_a V_{sm}}$ \\
\hline
$c_{15} = \dfrac{1}{\eta_{cp}}$ 
& $c_{16} = k_{ca,in}$ \\
\hline
$c_{17} = k_{ca,out}$ 
& $c_{18} = \eta_{cm} k_t$ \\
\hline
$c_{19} =  \dfrac{k_{ca,in} x_{O_2}}{(1 + \omega_{atm})}$ 
& $c_{20} = \dfrac{n M_{O_2}}{4F}$ \\
\hline
$c_{21} = \dfrac{\eta_{v-c} V_{cpr/tr} \rho_a}{2\pi} $ 
& $\times$ \\
\hline
\end{tabular}
\end{table}

\begin{remark}
In this paper, the model described by \eqref{eq_model} is not used to design the control law, but to emulate the behavior of the PEMFC air-feed system. \hfill $\square$
\end{remark}

\color{black}
\subsection{Control objective} \label{ssec_PEMFC_Control}
Using measurements $y_1$ and $y_2$ given by 
\begin{subequations}
    \begin{align}
        y_1 & = x_1 + x_2 + c_2 + w_1,\\
        y_2 & = x_4 + w_2,
    \end{align}
\end{subequations}
where $w_1$ and $w_2$ are independent zero mean Gaussian white noises.

The objective is to regulate the excess ratio between the oxygen flow rate entering the PEMFC and the oxygen flow rate consumed by the fuel cell due/from to the electrochemical reaction at a desired value. This excess ratio between flow rates is known as ``oxygen stoichiometry ($\lambda_{O_2}$)” and is defined by: 
\begin{equation}
 \lambda_{O_2} = \frac{c_{19}(y_2-y_1)}{c_{20}\xi}
    \label{eq_lambda}
\end{equation}
where $c_{19}$ and $c_{20}$ are known constants.
 
The desired oxygen stoichiometry ($\lambda_{O_2}^{\star}$), i.e. the setpoint, can be set as follows $ 2 \leqslant \lambda_{O_2}^{\star} \leqslant 2.5$  to ensure sufficient oxygen flow for the electrochemical reaction. It is important to note that a value $\lambda_{O_2} \leqslant 1$ can lead to a starvation fault which can generate some irreversible damages to the PEMFC stack. Furthermore, a value of $\lambda_{O_2} > 2.5$ may lead to an evacuation of water generated by the PEMFC, which causes the membrane drying and increases the compressor's energy consumption.  In the literature, there exist two main ways to define the desired oxygen stoichiometry $\lambda_{O_2}^{\star}$
\begin{enumerate}
    \item $\lambda_{O_2}^{\star}$ is chosen as a constant $ 2 \leqslant \lambda_{O_2}^{\star} \leqslant 2.5$  despite  variations of $\xi$ (see \cite{WWZX:21} and \cite{APJBCYD:22} ),
    \item $\lambda_{O_2}^{\star}$ is defined as a polynomial function of the stack current and given by (see \cite{MLJW:13} and \cite{MZGCS:20})
    \begin{multline}
    \lambda_{O_2}^{\star}  = 5 \times10^{-8}\xi^{3} - 2.87\times10^{-5}\xi^{2}\\+2.23\times10^{-3}\xi +2.5.
    \label{eq_stoech_desir}
\end{multline}
\end{enumerate}

\section{SIMULATIONS RESULTS} \label{sec_Sim}

\subsection{Scenarios}\label{ssec_scenarii}
Two scenarios are considered  for the simulations:
\begin{enumerate}  
    \item the \st{1} is devoted to regulating the oxygen stoichiometry $\lambda_{O_2}$ at a fixed value $\lambda_{O_2}^{\star} =2.2$,   
    \item the \nd{2} scenario is dedicated to regulating the oxygen stoichiometry $\lambda_{O_2}$ at a variable  value of $\lambda_{O_2}^{\star}$ which is given by \eqref{eq_stoech_desir}.
\end{enumerate}

\begin{figure}[h!]
    \centering
    \includegraphics[width=1.0\columnwidth]{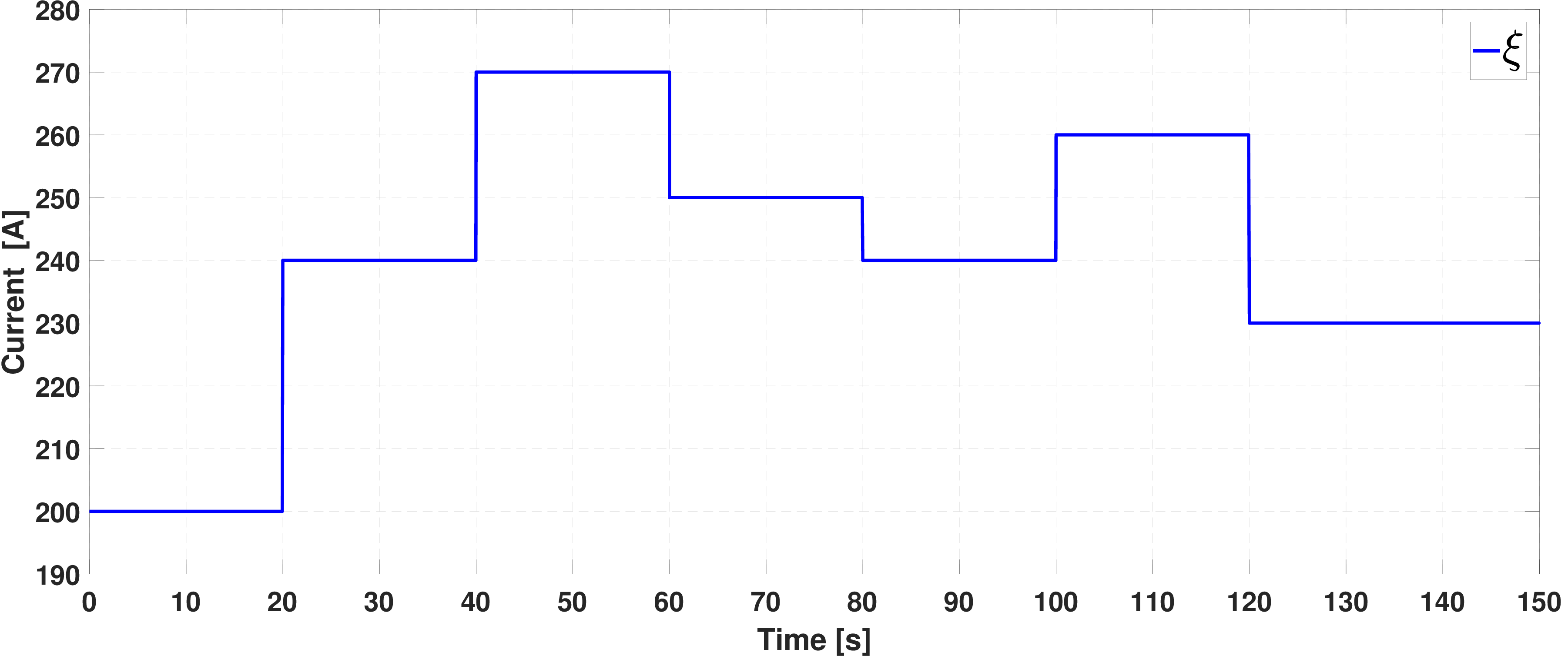}
    \caption{\st{1} current profile}
    \label{fig_current_not_hard}
\end{figure}

\begin{figure}[h!]
    \centering
    \includegraphics[width=1.0\columnwidth]{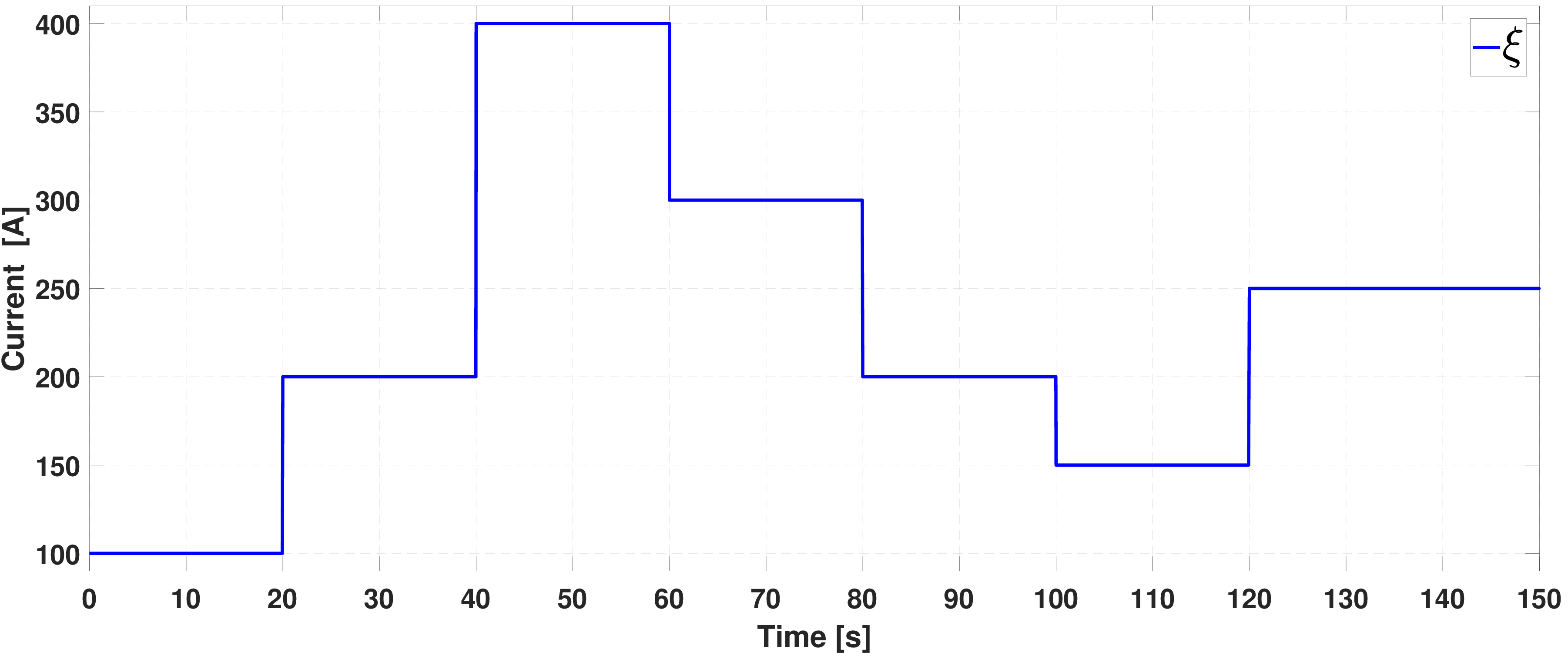}
    \caption{\nd{2} current profile}
    \label{fig_current_hard}
\end{figure}

For these two scenarios, two different current profiles of $\xi$ are considered. In the \st{1} current profile, the variations of $\xi$ can be regarded as small (see Fig. \ref{fig_current_not_hard}), while the \nd{2} current profile, which has significant variations of $\xi$, is taken into account (see Fig. \ref{fig_current_hard}).

In order to analyze the robustness of the \emph{iP} controller, two cases are considered:   
\begin{itemize}
    \item the physical parameters of PEMFC system in \eqref{eq_model} given in table 2 in \cite{MLJW:13} are considered for nominal case, 
    \item some physical parameters of PEMFC system in \eqref{eq_model} given in table 2 in \cite{MLJW:13} are modified and given in Table \ref{tab_uncer} for uncertain case.  
\end{itemize}

\begin{table}[h!]
\caption{Parameters uncertainties}
\label{tab_uncer}
\centering
\begin{tabular}{|c||c||c|}
\hline
Symbol & Parameter & Value\\
\hline
$f$ & Motor friction & $+20$\% \\
\hline
$k_t$ & Motor constant & $-$5\% \\
\hline
$\eta_{cp}$ & Compressor efficiency & $-$10\% \\
\hline
$\eta_{cm}$ & Motor mechanical efficiency & $-$20\% \\
\hline
$k_{ca,out}$ & Cathode outlet constant & $+$10\% \\
\hline
$T_{atm}$ & Atmospheric temperature & $+$10\% \\
\hline
$V_{ca}$ & Cathode volume & $+$10\% \\
\hline
$V_{sm}$ & Supply manifold volume & $-$10\% \\
\hline
$T_{fc}$ & Stack temperature, °C &  $+$12\%\\
\hline
\end{tabular}
\end{table}

The following notations are retained: superscript $n$ for the nominal case and superscript $u$ for the uncertain one.

\subsection{Constant desired oxygen stoichiometry}\label{ssec_sim_constant}

\subsubsection{\st{1} current profile}\label{sssec_sim_constant_first}

\begin{figure*}
\centering
\begin{subfigure}{0.49\textwidth}
    \includegraphics[width=\textwidth]{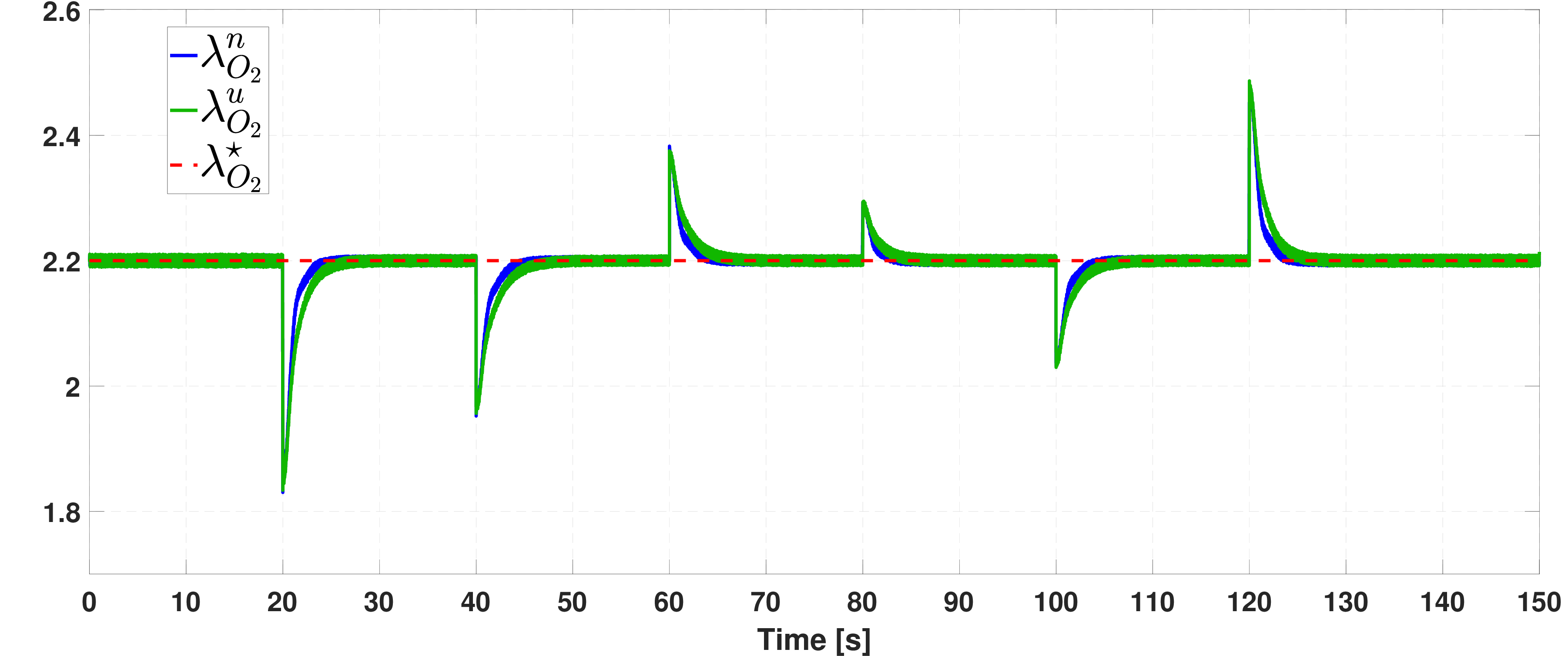}
    \caption{Constant $\lambda_{O_2}$ for \st{1} current profile}
    \label{fig_not_hard_fixe_lambda}
\end{subfigure}
\begin{subfigure}{0.49\textwidth}
    \includegraphics[width=\textwidth]{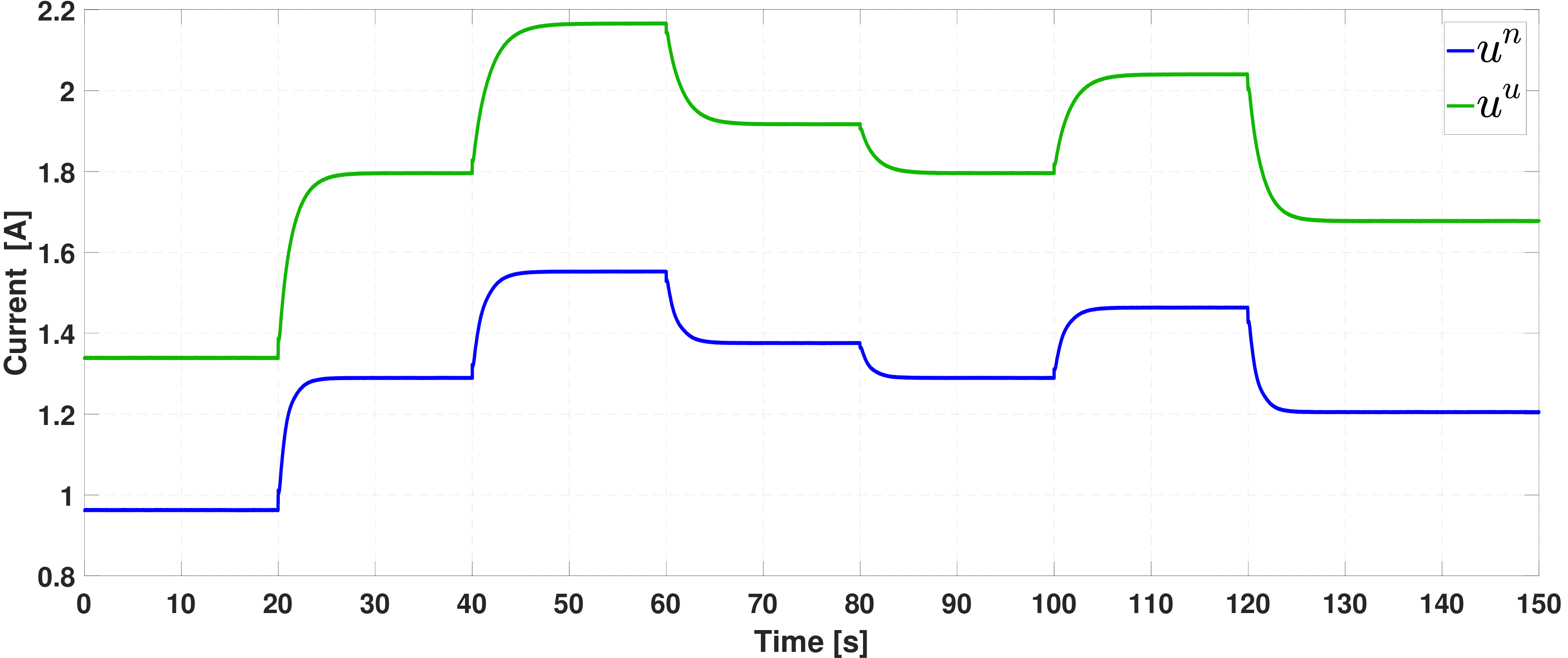}
    \caption{$u$ for \st{1} current profile}
    \label{fig_not_hard_fixe_u}
\end{subfigure}
\begin{subfigure}{0.49\textwidth}
    \includegraphics[width=\textwidth]{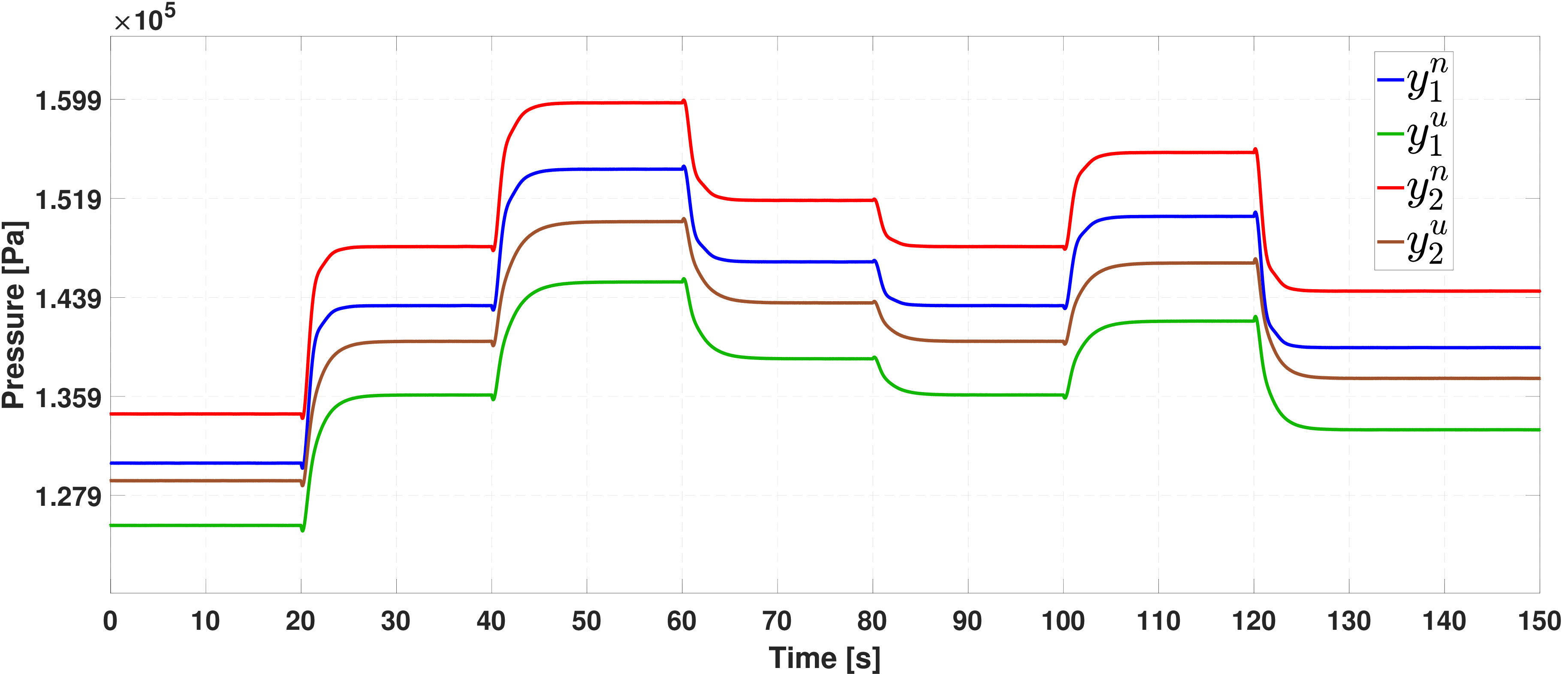}
    \caption{$y_1$ and $y_2$ for \st{1} current profile}
    \label{fig_not_hard_fixe_y}
\end{subfigure}
\caption{Constant desired oxygen stoichiometry $\lambda_{O_2}^{\star}$  with and without uncertainties: \st{1} current profile}
\label{fig_not_hard_fixe}
\end{figure*}

Fig. \ref{fig_not_hard_fixe} shows the control of oxygen stoichiometry when \st{1} profile shown in Fig. \ref{fig_current_not_hard}  is applied. $\lambda_{O_2}$ is restored to $\lambda_{O_2}^{\star}$ in 5\,s for all six current changes for the nominal case, see Fig. \ref{fig_not_hard_fixe_lambda}.  For the uncertain case, the \emph{iP} controller performs very well with regard to these uncertainties in the PEMFC system parameters, ensuring a restoration time of $\lambda_{O_2}$ to $\lambda_{O_2}^{\star}$ in 6.5\,s. 

The differences between the control inputs $u^{n}$ and $u^{u}$ shown in Fig. \ref{fig_not_hard_fixe_u} are due to all the uncertainties in the PEMFC parameters and confirms that the controller is capable of tolerating these uncertainties. However, the differences between $y_1^{n}$, $y_1^{u}$ and $y_2^{n}$, $y_2^{u}$, see Fig. \ref{fig_not_hard_fixe_y}, are only due to uncertainties associated with the stack parameters ($k_{ca,in}$, $k_{ca,out}$, $T_ {atm}$, $V_{ca}$, $V_{sm}$ and $T_ {fc}$), but not to uncertainties related to the compressor parameters ($f$, $k_t$, $\eta_{cp}$ and $\eta_{cm}$).


\subsubsection{\nd{2} current profile}\label{sssec_sim_constant_second}

\begin{figure*}[h!]
\centering
\begin{subfigure}{0.49\textwidth}
    \includegraphics[width=\textwidth]{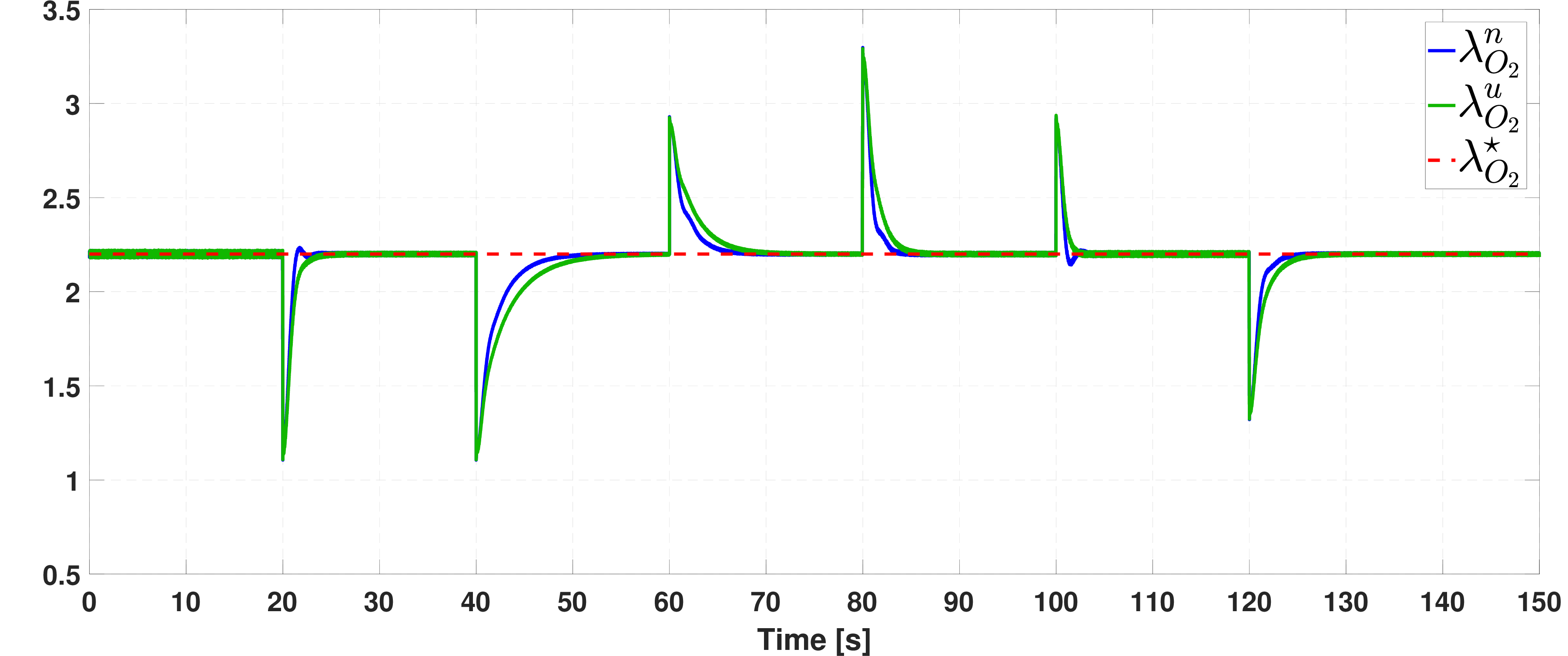}
    \caption{Constant $\lambda_{O_2}$ for \nd{2} current profile}
    \label{fig_hard_fixe_lamda}
\end{subfigure}
\begin{subfigure}{0.49\textwidth}
    \includegraphics[width=\textwidth]{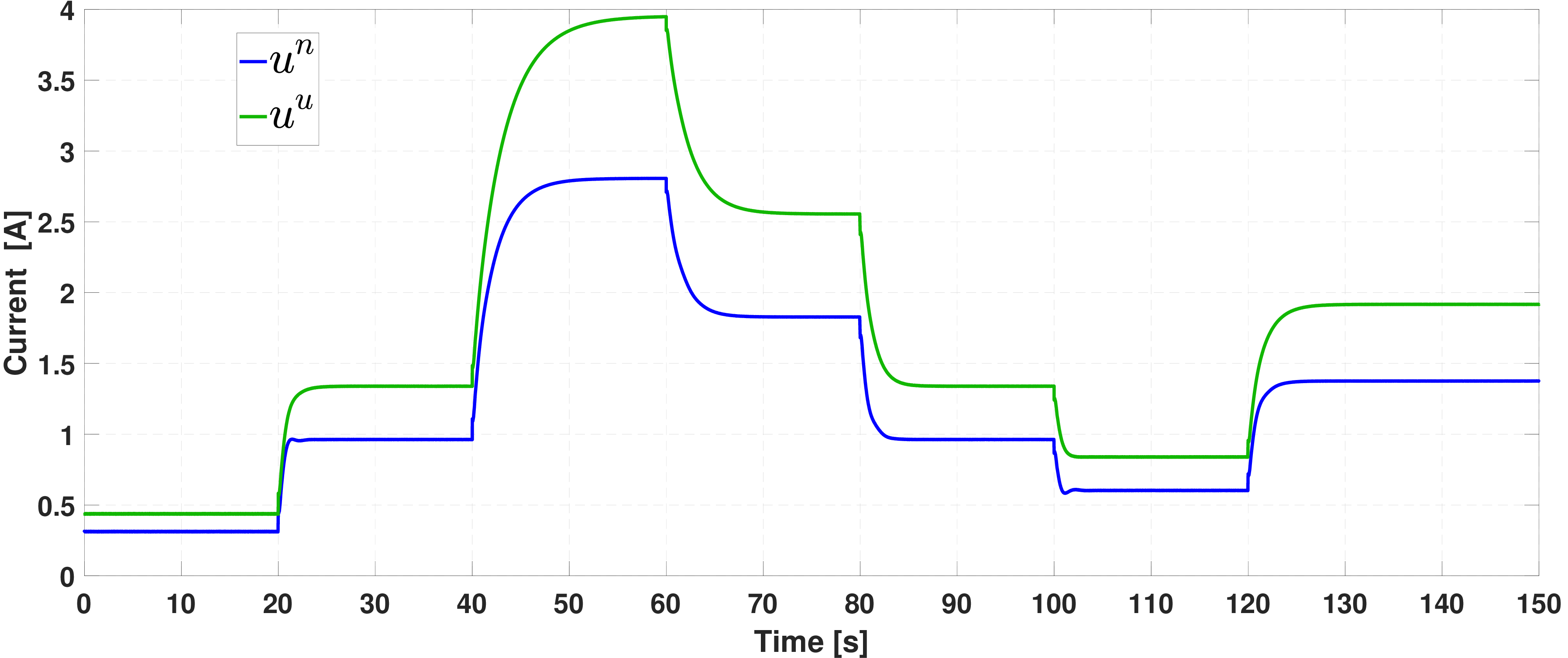}
    \caption{$u$ for \nd{2} current profile}
    \label{fig_hard_fixe_u}
\end{subfigure}
\begin{subfigure}{0.49\textwidth}
    \includegraphics[width=\textwidth]{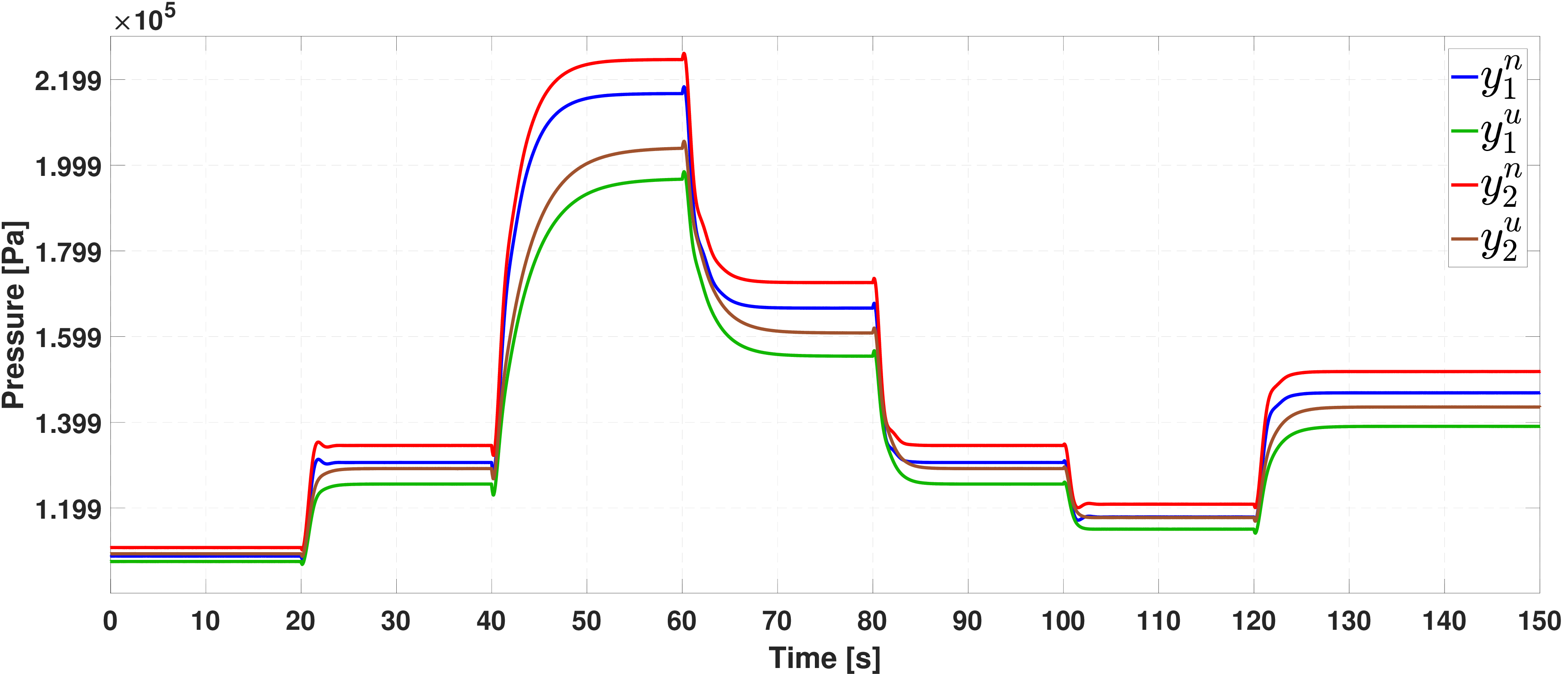}
    \caption{$y_1$ and $y_2$ for \nd{2} current profile}
    \label{fig_hard_fixe_y}
\end{subfigure}
\caption{Constant desired oxygen stoichiometry $\lambda_{O_2}^{\star}$ with and without uncertainties: \nd{2} current profile}
\label{fig_hard_fixe}
\end{figure*}

Fig. \ref{fig_hard_fixe} shows the oxygen stoichiometry control when the \nd{2} profile is applied. This profile presents higher current variations than the \st{1} one. For the nominal case, the \emph{iP} controller resets $\lambda_{O_2}$ to $\lambda_{O_2}^{\star}$ between 2\,s and 10\,s depending on the magnitude of the variation in current $\xi$. With regard to the uncertain case, it should be noted that there is no significant difference; the restoration time is 1.5\,s longer than in the nominal case as shown in Fig. \ref{fig_hard_fixe_lamda}. The same comments can be made for Fig. \ref{fig_hard_fixe_u} and Fig. \ref{fig_hard_fixe_y} as for Fig. \ref{fig_not_hard_fixe_u} and  Fig. \ref{fig_not_hard_fixe_y}. 

The simulation results show that the \emph{iP} controller is robust with respect to parameter uncertainties and ensures satisfactory closed-loop behaviors for the two scenarii with constant desired oxygen stoichiometry.


\subsection{Variable desired oxygen stoichiometry} \label{ssec_sim_variable}
In this Section, the desired oxygen stoichiometry $\lambda_{O_2}^{\star}$ is  variable, so the control objective is to regulate $\lambda_{O_2}$ to $\lambda_{O_2}^{\star}$ even with  PEMFC parameter uncertainties under the two above-mentioned current profiles.

\subsubsection{\st{1} current profile}\label{sssec_sim_variable_first}

\begin{figure*}[h!]
\centering
\begin{subfigure}{0.49\textwidth}
    \includegraphics[width=\textwidth]{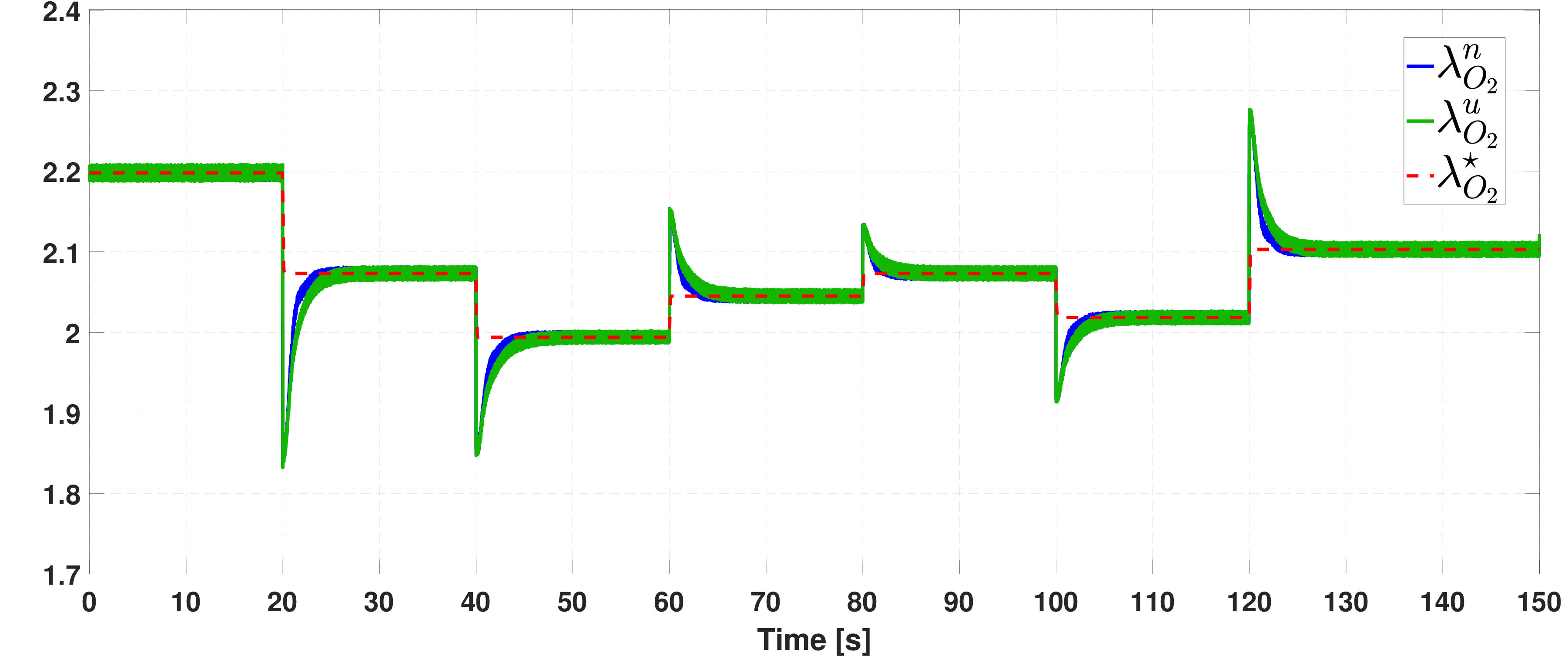}
    \caption{Variable $\lambda_{O_2}$ for \st{1} current profile}
    \label{fig_not_hard_var_lambda}
\end{subfigure}
\begin{subfigure}{0.49\textwidth}
    \includegraphics[width=\textwidth]{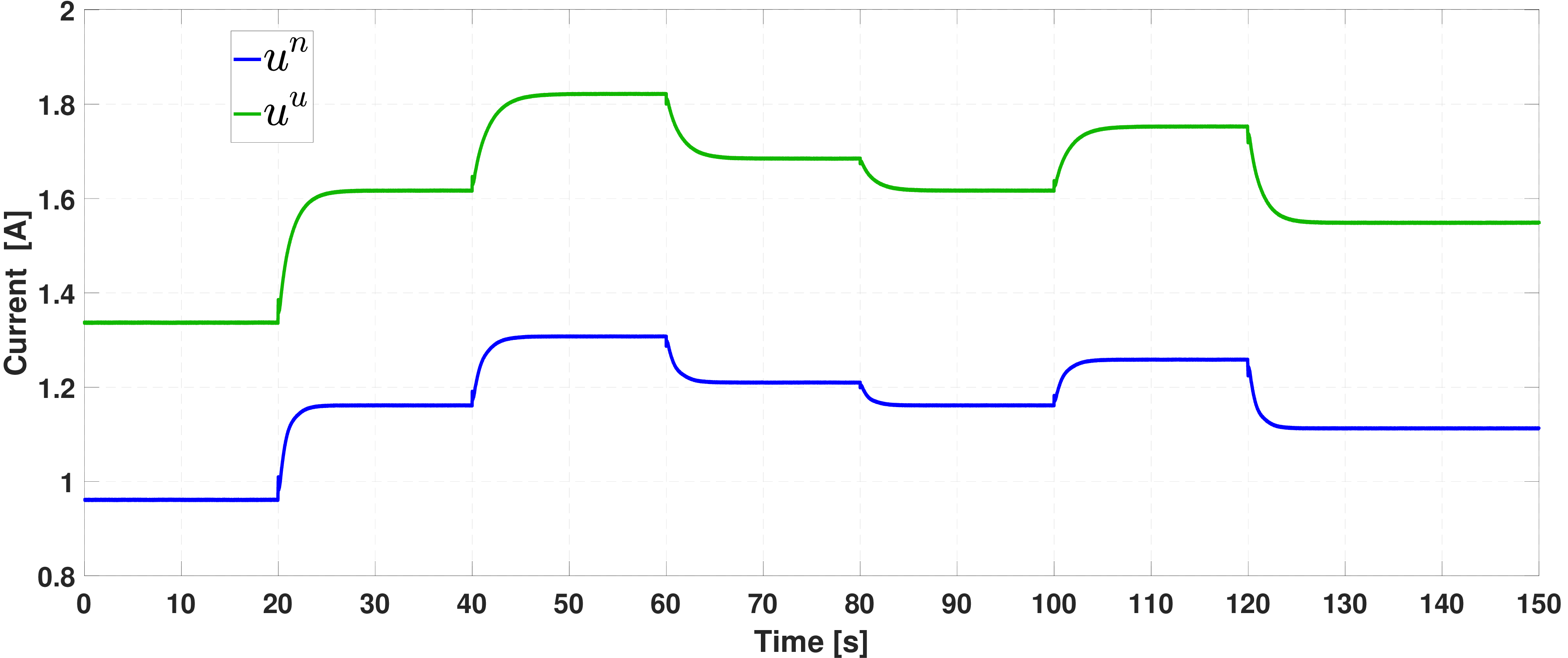}
    \caption{$u$ for \st{1} current profile}
    \label{fig_not_hard_var_u}
\end{subfigure}
\begin{subfigure}{0.49\textwidth}
    \includegraphics[width=\textwidth]{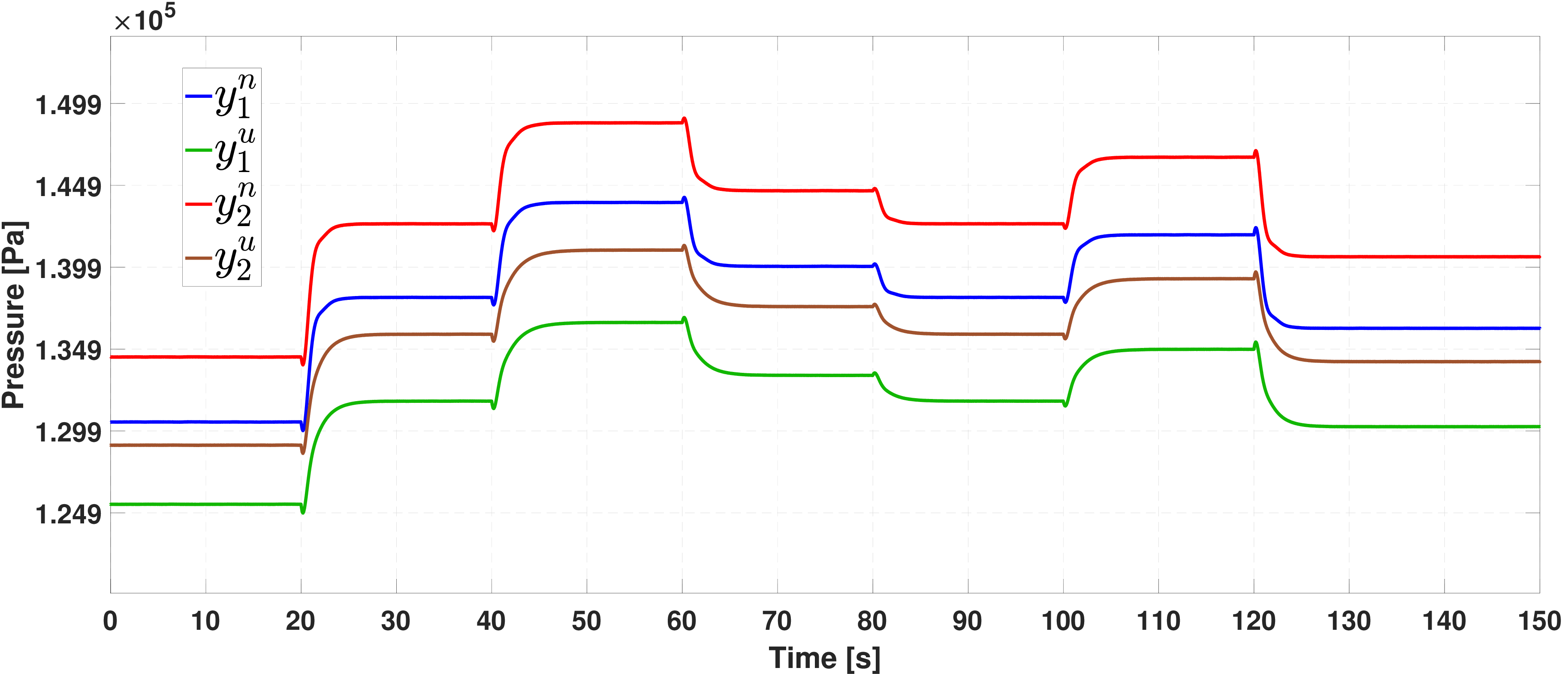}
    \caption{$y_1$ and $y_2$ for \st{1} current profile}
    \label{fig_not_hard_var_y}
\end{subfigure}
\caption{Variable desired oxygen stoichiometry $\lambda_{O_2}^{\star}$  with and without uncertainties: \st{1} current profile}
\label{fig_not_hard_var}
\end{figure*}

For the \st{1} profile $2 \leq \lambda_{O_2}^{\star} \leq 2.2$. The \emph{iP} controller ensures both disturbance rejection and $\lambda_{O_2}^{\star}$ tracking   in 5\,s for the nominal case and in 6\,s for the uncertain case, as shown in Figure \ref{fig_not_hard_var_lambda}. The robustness of \emph{iP} controller is achieved since the responses are close in nominal and uncertain cases in Figure \ref{fig_not_hard_var_lambda}. The control input and the measurements for both nominal  and uncertain cases are shows in Fig. \ref{fig_not_hard_var_u} and Fig. \ref{fig_not_hard_var_y}, respectively. For both constant and variable desired oxygen stoichiometry, the transient behavior is closed in the two cases (see the settling times and the peaks in Fig. \ref{fig_not_hard_fixe_lambda} and Fig. \ref{fig_not_hard_var_lambda}). However, the difference between the control inputs and the measurements in both cases is mainly explained by the difference in the setpoints (see Fig. \ref{fig_not_hard_fixe_u}, Fig. \ref{fig_not_hard_fixe_y}, Fig. \ref{fig_not_hard_var_u}, and Fig. \ref{fig_not_hard_var_y}).


\subsubsection{\nd{2} current profile}\label{sssec_sim_variable_second}

\begin{figure*}
\centering
\begin{subfigure}{0.49\textwidth}
    \includegraphics[width=\textwidth]{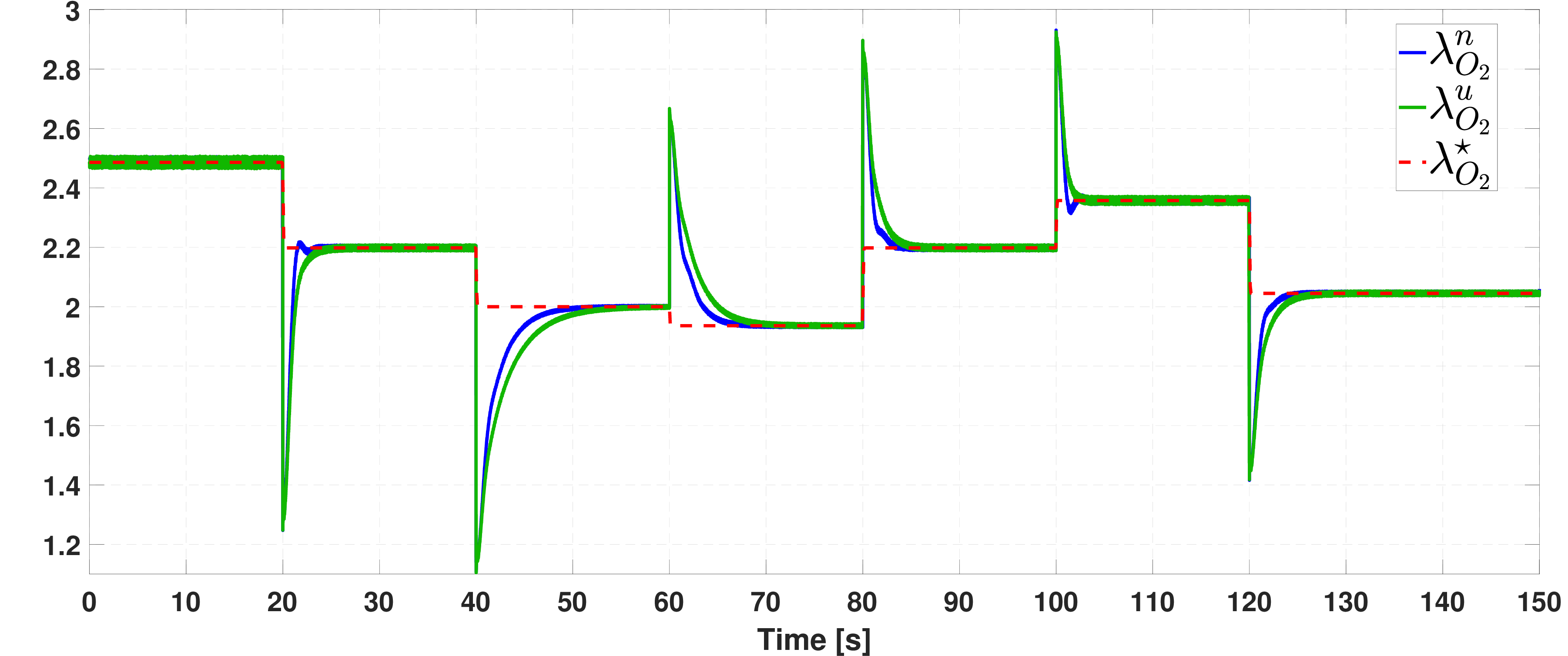}
    \caption{Variable $\lambda_{O_2}$ for \nd{2} current profile}
    \label{fig_hard_var_lambda}
\end{subfigure}
\begin{subfigure}{0.49\textwidth}
    \includegraphics[width=\textwidth]{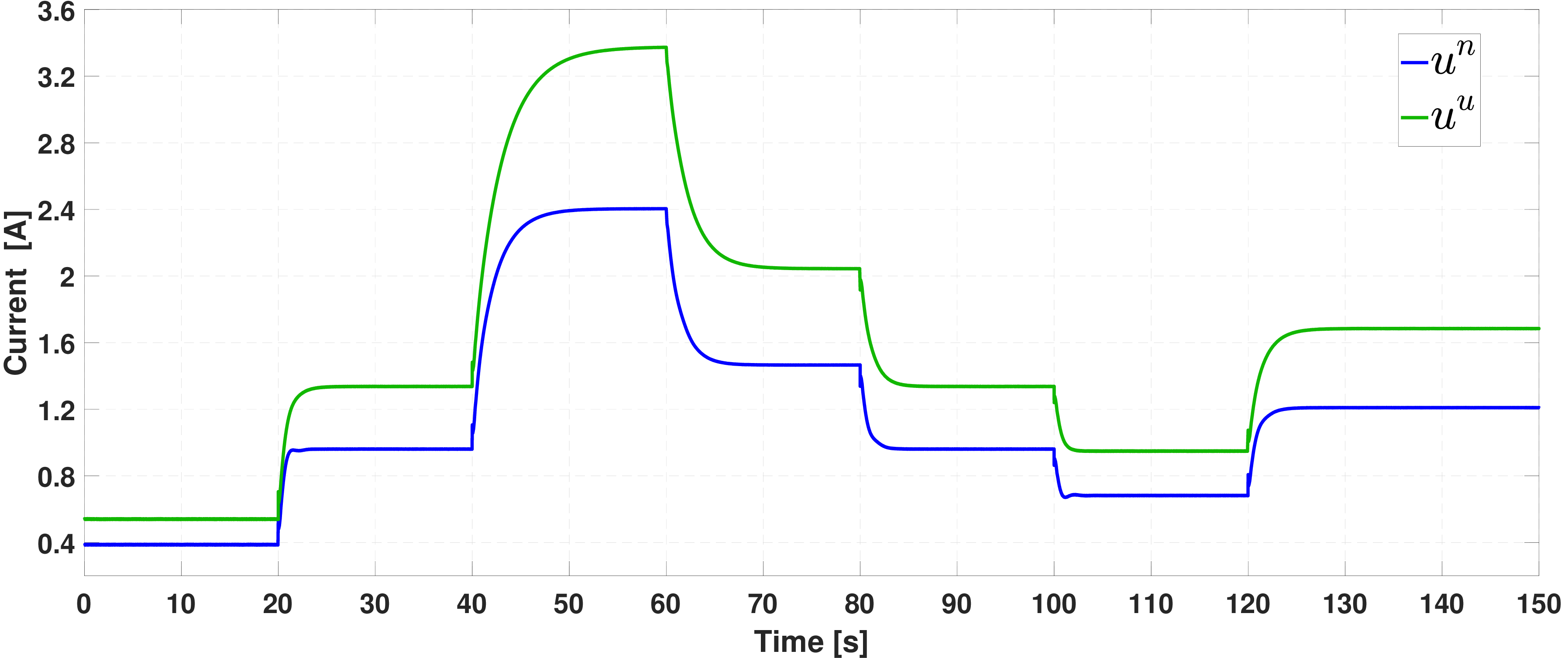}
    \caption{$u$ for \nd{2} current profile}
    \label{fig_hard_var_u}
\end{subfigure}
\begin{subfigure}{0.49\textwidth}
    \includegraphics[width=\textwidth]{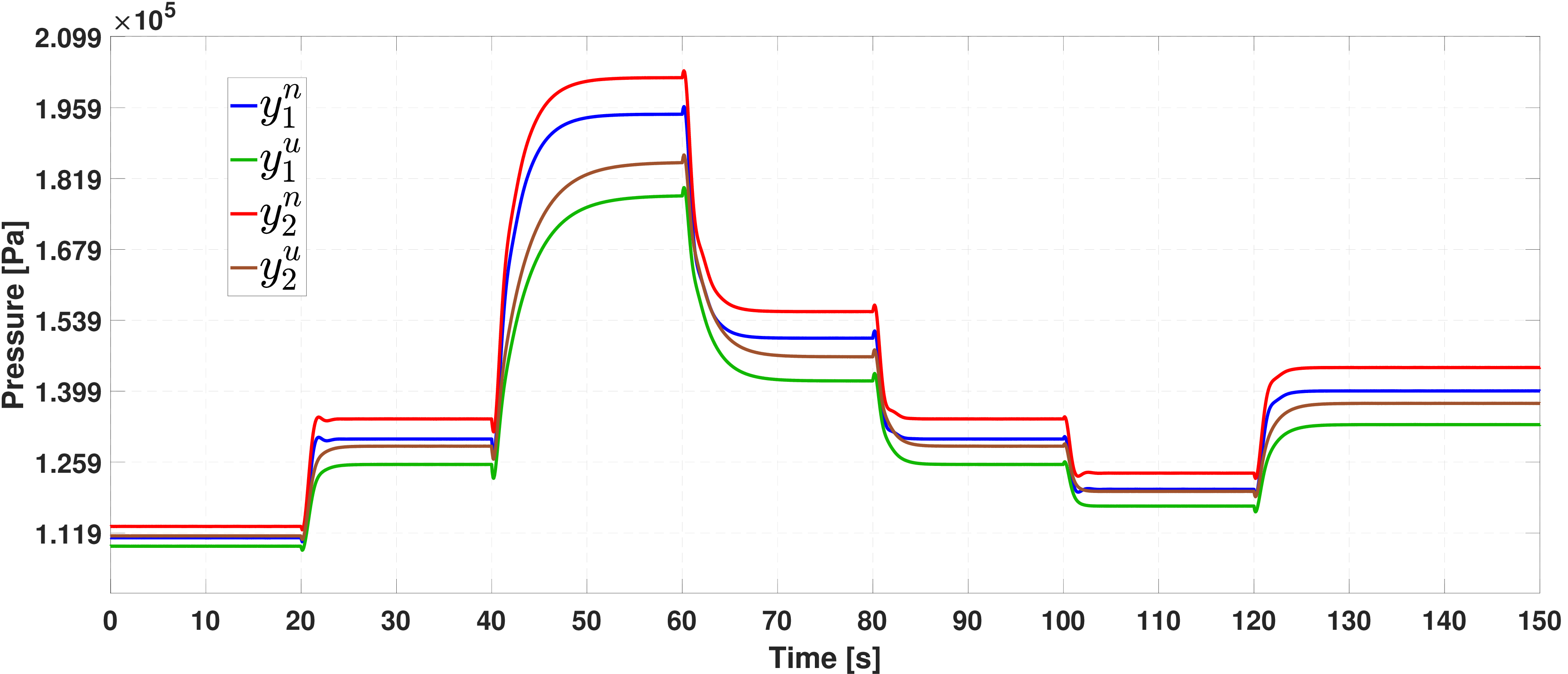}
    \caption{$y_1$ and $y_2$ for \nd{2} current profile}
    \label{fig_hard_var_y}
\end{subfigure}
\caption{Variable desired oxygen stoichiometry $\lambda_{O_2}^{\star}$  with and without uncertainties: \nd{2} current profile}
\label{fig_hard_var}
\end{figure*}

For the \nd{2} profile $1.9 \leq \lambda_{O_2}^{\star} \leq 2.5$. Fig. \ref{fig_hard_var} shows the oxygen stoichiometry control when the \nd{2} profile is applied. Like in the \st{1} scenario, the \emph{iP} controller resets $\lambda_{O_2}$ to $\lambda_{O_2}^{\star}$ between 2\,s and 10\,s depending on the magnitude of the variation in current $\xi$ for the nominal case, and there is no significant difference for the uncertain case (the tracking time is  1.5\,s longer than in the nominal case as shown in Fig. \ref{fig_hard_var_lambda}). The same comments can be made for Fig. \ref{fig_hard_var_u} and Fig. \ref{fig_hard_var_y} as for Fig. \ref{fig_not_hard_var_u} and  Fig. \ref{fig_not_hard_var_y}. 

The above simulations illustrate the fact that the \emph{iP} controller ensures good reference tracking and robustness against parameter uncertainties for both scenarios with a variable desired oxygen stoichiometry.



\section{CONCLUSIONS}\label{sec_Conc}
The results obtained with the \emph{iP} controller are satisfactory in simulation with a model widely used in the literature and allow us to proceed to the next step: implementing the \emph{iP} controller in a real test bench. This real-time implementation allows us to compare the proposed approach with those mentioned in the Introduction. 
This comparison will be carried out in terms of both the performance and robustness of the control laws and the computational burden in order to propose effective control strategies for this complex system. 

Another perspective of this work is to address PEM fuel cell energy management \cite{BTDYMB:25} and fault-tolerant  control and diagnosis  using model-free approaches \cite{AZJP:25,AJZ:25}, particularly to deal with sensor faults that affect $\lambda_{O_2}$ in \eqref{eq_lambda}.

\color{black}









\bibliography{MZ-1} 
\end{document}